\documentstyle[aps]{revtex}
\begin{document}
\draft
\preprint{SISSA 3/99/EP}
\title{Detection of Neutral MSSM Higgs Bosons at LEP-II and NLC}
\author{U. Cotti $^{a,}$ \thanks{Permanent address: Centro de Investigaci\'on
y de Estudios Avanzados del Instituto Polit\'ecnico Nacional, Apartado Postal
14-740, 07000 M\'exico, D.F. M\'exico}, A. Guti\'errez-Rodr\'{\i}guez $^{b}$,
A. Rosado $^{b}$, and O. A. Sampayo $^{c}$}

\address{(a) SISSA-ISAS, via Beirut 2-4, 34013 Trieste,
Italy.}

\address{(b) Instituto de F\'{\i}sica , Universidad Aut\'onoma de Puebla,\\
Apartado Postal J-48, Col. San Manuel, Puebla, Pue. 72570, M\'exico.}

\address{(c) Departamento de F\'{\i}sica, Universidad Nacional de Mar de Plata,\\
Funes 3350, (7600) Mar de Plata,  Argentina.}

\date{\today}
\maketitle
\begin{abstract}
We study the possibility of detecting the neutral Higgs bosons predicted in
the Minimal Supersymmetric Standard Model $(h^0, H^0, A^0)$, with the
reactions $e^{+}e^{-}\rightarrow b\bar b h^0 (H^0, A^0)$, using the
helicity formalism. We analyze the region of parameter space
$(m_{A^0}-\tan\beta)$ where $h^0(H^0, A^0)$ could be detected in the
limit when $\tan\beta$ is large. The numerical computation is done for the
energy which is expected to be available at LEP-II ($\sqrt{s}=200$ $GeV$)
and for a possible Next Linear $e^{+}e^{-}$ Collider ($\sqrt{s}=500$ $GeV$).
\end{abstract}
\pacs{PACS: 14.80.Cp, 12.60.Jv}

\section{Introduction}
The Higgs sector is one of the most important areas of the Standard Model
(SM) that has not been yet tested \cite{1}. The SM contain only one neutral 
Higgs boson and although its detection would give more validity to the SM, 
there are some theoretical problems that suggest the need for new physics. 
One of the more attractive extensions of the SM is Supersymmetry (SUSY)
\cite{2}, mainly because of its capacity to solve the naturalness and 
hierarchy problems while maintaining the Higgs bosons elementary.

The minimal supersymmetric extension of the Standard Model (MSSM) doubles the
spectrum of particles of the SM and the new free parameters obey simple
relations. The scalar sector of the MSSM \cite{3} requires two Higgs doublets,
thus the remaining scalar spectrum contains the  following physical states:
two CP-even Higgs scalar ($h^0$ and $H^0$) with $m_{h^0}\leq m_{H^0}$,
one CP-odd Higgs scalar ($A^0$) and a charged Higgs pair ($H^{\pm}$), whose
detection would be a clear signal of new physics. The Higgs sector is
specified at tree-level by fixing two parameters, which can be chosen as the
mass of the pseudoscalar $m_{A^0}$ and the ratio of vacuum expectation
values of the two doublets $\tan \beta = v_{2}/v_{1}$, then the mass
$m_{h^0}$, $m_{H^0}$ and $m_{H^{\pm}}$ and the mixing angle of the neutral
Higgs sector $\alpha$ can be fixed. However, since radiative corrections
produce substantial effects on the predictions of the model \cite{4}, it is
necessary to specify also the squark masses, which are assumed to be
degenerated. In this paper, we focus on the phenomenology of the neutral
CP-even and CP-odd scalar ($h^0, H^0, A^0$).

The search for these scalars has begun at LEP, and current low energy bound
on their masses gives $m_{h^0}$, $m_{A^0}$ $>$ 75 $GeV$ for $\tan\beta$
$>$ 1 \cite{5}.

At $e^{+}e^{-}$ colliders the signals for Higgs bosons are relatively clean
and the opportunities for discovery and detailed study will be excellent. The
most important processes for the production and detection of the neutral
Higgs bosons, $h^0$, $H^0$ and $A^0$, are: $e^{+}e^{-}\rightarrow
Z^{*}\rightarrow h^0, H^0+Z^0$, $e^{+}e^{-}\rightarrow Z^{*} \rightarrow
h^0, H^0+A^0$, and $e^{+}e^{-}\rightarrow \nu \bar \nu + W^{+*}W^{-*}
\rightarrow \nu \bar \nu+h^0, H^0$ (the later is conventionally referred
to as $WW$ fusions); precise cross section formulae appear in Ref. \cite{6}.
There is considerable complementarity among the first four processes, and the
$WW$ fusion processes are also complementary to one another and to the first
four. If $m_{A^0}>\!\!\!\!\!\! _{\sim} m_{Z^0}$, so that $\cos^{2}(\beta -
\alpha)$ is small, then $Z^{*} \rightarrow h^0Z^0$, $Z^{*} \rightarrow
H^0A^0$, and $WW\rightarrow h^0$ are all maximal, and the other three
small.

In particular, an $h^0$ with $m_{h^0}\sim m_{Z^0}$ could be seen at LEP-II
provided that $\sqrt{s}>\!\!\!\!\!\! _{\sim}\hspace*{2mm}200$ $GeV$ with
${\cal L}\sim 500$ $pb^{-}$ \cite{7} can be achieved, and that efficient
$b$-tagging is possible \cite{8}. But, if the $h^0$ were not discovered at
this energy, pushing slowly to $\sqrt{s} = 240$ $GeV$ would rapidly open up
the possibility for $h^0$ detection in the regions of parameters space
corresponding to the higher values of $M_{\stackrel {\sim}t}$ and
$\tan \beta$. Of course, the ability of LEP-II to detect $h^0$ is also
strongly dependent upon the actual $\sqrt{s}$ that can be achieved and upon
the unknown value of $M_{\stackrel {\sim}t}$ (and to a lesser degree on the
parameters that control squark mixing which in this paper will be neglected).
Since radiative corrections make possible to have $m_{h^0} > m_{Z^0}$, it
seems that hadron colliders will be also required to test fully the Higgs
sector of the MSSM \cite{9}.

The $Z^0h^0$ production cross section contains an overall factor $\sin^{2}(\beta-\alpha)$
which suppress it in certain parameter regions (with $m_{A^0} < 100$ $GeV$
and $\tan\beta$ large); fortunately the $A^0h^0$ production cross section contains
the complementary factor $\cos^{2}(\beta-\alpha)$. Hence the $Z^0h^0$ and $A^0h^0$
channels together are well suited to cover all regions in the $(m_{A^0}-\tan\beta)$
plane, provided that the $c.m.$ energy is high enough for $Z^0h^0$ to be produced through
the whole $m_{h^0}$ mass range, and that an adequate event rate can be achieved. 
These conditions are already shown to be satisfied \cite{10} for $\sqrt{s}= 500$ $GeV$
with assumed luminosity $10$ $fb^{-1}$, like is expected to be the case of the Next Linear $e^{+}e^{-}$
Collider (NLC).

In previous studies, the two body processes $e^{+}e^{-}\rightarrow h^0(H^0)+Z^0$
and $e^{+}e^{-}\rightarrow h^0(H^0)+A^0$ have been evaluated \cite{6} extensively.
However, the inclusion of three-body process $e^{+}e^{-}\rightarrow h^0(H^0)+b\bar b$
and $e^{+}e^{-}\rightarrow A^0+b\bar b$ at LEP-II and NLC energies is
necessary in order to know its impact on the two body mode processes and
also to search for new relations that could have a cleaner signature of the Higgs
bosons production.

In the present paper we study the production of SUSY Higgs bosons at $e^{+}e^{-}$ colliders.
We are interested in finding regions that could allow the detection of the SUSY
Higgs bosons for the set parameter space $(m_{A^0}-\tan\beta)$. We shall discuss
the neutral Higgs bosons production $b\bar b h^0(H^0, A^0)$ in the energy range 
of LEP-II and NLC for large values of the parameter $\tan\beta$,
where one expects to have a high production. Since the coupling
$h^0b\bar b$ is proportional to $\sin\alpha/\cos\beta$, the cross section
will receive a large enhancement factor when $\tan\beta$
is large. Similar situation occurs for $H^0$, whose coupling with $b\bar b$
is proportional to $\cos\alpha/\cos\beta$. The coupling of $A^0$ with $b\bar b$ is
directly proportional to $\tan\beta$, thus the amplitude will always grow with $\tan\beta$. We
consider the complete set of Feynman diagrams at tree-level and use the helicity
formalism \cite{11,12,13,14,15,16,17} for the evaluation of the amplitudes.
The results obtained for the
3-body processes are compared with the dominant mode 2-body reactions for the 
plane  $(m_{A^0}-\tan\beta)$. Succinctly, our aim in this work is to analyze
how much the results of the Bjorken Mechanism (Fig. 1.4) would be enhanced by
the contribution from the diagrams depicted in Figs. 1.1, 1.2, 1.3, 1.5 and
1.6, in which the SUSY Higgs boson is radiated by a $b(\bar b)$ quark.

Recently, it has been shown that for large values of $\tan\beta$ detection
of SUSY Higgs bosons is possible at FNAL and LHC \cite{18}. In the papers
cited in Ref.[18] the authors calculated the corresponding 3-body diagrams
for hadron collisions. They pointed out the importance of a large bottom
Yukawa coupling at hadron colliders and showed that the Tevatron collider
may be a good place for detecting SUSY Higgs bosons. In the case of the
hadron colliders the three body diagrams come from gluon fusion and this
fact makes the contribution from these diagrams more important, due to the
gluon abundance inside the hadrons. The advantage
for the case of $e^{+} e^{-}$ colliders is that the signals of the processes
are cleaner.

This paper is organized as follows. We present in Sect. II the relevant details of the 
calculations. Sect. III contains the results for the process $e^{+}e^{-}\rightarrow b\bar bh^0(H^0,A^0)$
at LEP-II and NLC. Finally, Sect. IV contains our conclusions.

\section{Helicity Amplitude for Neutral Higgs Bosons Production}

When the number of Feynman diagrams is increased, the calculation of the amplitude
is a rather unpleasant task. Some algebraic forms \cite{19} can be used in it to
avoid manual calculation, but sometimes the lengthy printed output from the computer
is overwhelming, and one can hardly find the required results from it. The CALKUL 
collaboration \cite{20} suggested the Helicity Amplitude Method (HAM) which can
simplify the calculation remarkably and hence make the manual calculation realistic.

In this section we discuss the evaluation of the amplitudes at tree-level, for 
$e^{+}e^{-}\rightarrow b\bar b h^0(H^0, A^0)$ using the HAM \cite{11,12,13,14,15,16,17}.
This method is a powerfull technique for computing helicity amplitudes for multiparticle
processes involving massles spin-1/2 and spin-1 particles. Generalization of this method 
that incorporates massive spin-1/2 and spin-1 particles, are given in Ref. \cite{17}.
This algebra is easy to program and more efficient than computing the Dirac algebra.

A Higgs boson $h^0, H^0$ and $A^0$ can be produced in scattering $e^{+}e^{-}$ via
the following processes:

\begin{eqnarray}
e^{+}e^{-} &\rightarrow& b\bar b h^0,\\
e^{+}e^{-} &\rightarrow& b\bar b H^0,\\
e^{+}e^{-} &\rightarrow& b\bar b A^0.
\end{eqnarray}

The diagrams of Feynman, which contribute at tree-level to the different reaction
mechanisms are depicted in Figs. 1-2. Using the Feynman rules given by the
Minimal Supersymmetric Standard Model (MSSM), as are summarized in
Ref. \cite{6}, we can write the amplitudes for these reactions. For the
evaluation of the amplitudes we have used the spinor-helicity technique of
Xu, Zhang and Chang \cite{12} (denoted henceforth
by XZC), which is a modification of the technique developed by the CALKUL collaboration \cite{20}.
Following XZC, we introduce a very useful notation for the calculation of the processes (1)-(3).

\subsection{Cases $b\bar b h^0$ and $b\bar b H^0$}

Let us consider the process

\begin{equation}
e^{-}(p_{1}) + e^{+}(p_{2}) \rightarrow b(k_{2}) + \bar b(k_{3}) + h(k_{1}), \hspace{2mm} h = h^0,H^0
\end{equation}

\noindent in which the helicity amplitude is denoted by ${\cal M}(\lambda (e^{+}), \lambda (e^{-}), \lambda (b), \lambda (\bar b))$.
The Feynman diagrams for this process are shown in Fig. 1. From this figure it follows
that the amplitudes that correspond to each graph are:

\begin{eqnarray}
{\cal M}_1&=&-iC_{1}P_{Z^0b}(s_{13})P_{Z^0e}(s)T_{Z^0b}^{\mu}T_{Z^0e \mu},\nonumber\\
{\cal M}_2&=&-iC_{2}P_{Z^0b}(s_{12})P_{Z^0e}(s)T_{Z^0b}^{\mu}T_{Z^0e \mu},\nonumber\\
{\cal M}_3&=&-iC_{3}P_{A^0b}(s_{23})P_{Z^0e}(s)T^{\mu}_{A^0b}T_{Z^0e \mu},\\
{\cal M}_4&=&iC_{4}P_{Z^0b}(s_{23})P_{Z^0e}(s)T^{\mu}_{Z^0b}T_{Z^0e \mu},\nonumber\\
{\cal M}_5&=&-iC_{5}P_{\gamma b}(s_{13})P_{\gamma e}(s)T^{\mu}_{\gamma b}T_{\gamma e \mu},\nonumber\\
{\cal M}_6&=&-iC_{6}P_{\gamma b}(s_{12})P_{\gamma e}(s)T^{\mu}_{\gamma b}T_{\gamma e \mu},\nonumber
\end{eqnarray}

\noindent where

\begin{eqnarray}
C_{1}&=&\left\{ \begin{array}{ll}
               g^{3}m_{b}\sin \alpha /32M_{W}\cos ^{2}\theta _{W}\cos \beta,& \mbox{$h=h^0$} \\
               -g^{3}m_{b}\cos \alpha /32M_{W}\cos ^{2}\theta _{W}\cos \beta,& \mbox{$h=H^0$}
               \end{array}
      \right. \nonumber\\
C_{2}&=&\left\{ \begin{array}{ll}
               g^{3}m_{b}\sin \alpha /32M_{W}\cos ^{2}\theta _{W}\cos \beta,& \mbox{$h=h^0$} \\
               -g^{3}m_{b}\cos \alpha /32M_{W}\cos ^{2}\theta_ {W}\cos \beta,& \mbox{$h=H^0$}
               \end{array}
      \right. \nonumber\\
C_{3}&=&\left\{ \begin{array}{ll}
                g^{3}m_{b}\tan \beta \cos (\beta - \alpha)/16M_{W}\cos ^{2}\theta _{W},& \mbox{$h=h^0$}\\
                -g^{3}m_{b}\tan \beta \sin (\beta -\alpha)/16M_{W}\cos ^{2}\theta _{W},& \mbox{$h=H^0$}
                \end{array}
      \right. \nonumber\\
C_{4}&=&\left\{ \begin{array}{ll}
                g^{3}M_{Z^0}\sin (\beta - \alpha )/4\cos ^{3}\theta _{W},& \mbox{$h=h^0$} \\
                g^{3}M_{Z^0}\cos (\beta - \alpha )/4\cos ^{3}\theta _{W},& \mbox{$h=H^0$}
                \end{array}
      \right. \\
C_{5}&=&\left\{ \begin{array}{ll}
                gm_{b}e^{2}Q_{e}Q_{b}\sin \alpha /2M_{W}\cos \beta,& \mbox{$h=h^0$}\\
                -gm_{b}e^{2}Q_{e}Q_{b}\cos \alpha /2M_{W}\cos \beta, & \mbox{$h=H^0$}
                \end{array}
      \right. \nonumber\\
C_{6}&=&\left\{ \begin{array}{ll}
                gm_{b}e^{2}Q_{e}Q_{b}\sin \alpha /2M_{W}\cos \beta,& \mbox{$h=h^0$}\\
                -gm_{b}e^{2}Q_{e}Q_{b}\cos \alpha /2M_{W}\cos \beta,& \mbox{$h=H^0$}
                \end{array}
       \right.\nonumber
\end{eqnarray}

\noindent while the propagators are

\begin{eqnarray}
P_{Z^0b}(s_{13})&=&\frac{1}{(k_{1}+k_{3})^{2}},\nonumber\\
P_{Z^0e}(s)&=&\frac{1}{(s-M_{Z^0}^{2}-iM_{Z^0}\Gamma_{Z^0})},\nonumber\\
P_{Z^0b}(s_{12})&=&\frac{1}{(k_{1}+k_{2})^{2}},\nonumber\\
P_{hb}(s_{23})&=&\frac{1}{[(k_{2}+k_{3})^{2}-M_{h}^{2}-iM_{h}\Gamma _{h}]}, \mbox{$h=h^0,H^0, A^0$}\nonumber\\
P_{Z^0b}(s_{23})&=&\frac{1}{[(k_{2}+k_{3})^{2}-M_{Z^0}^{2}-iM_{Z^0}\Gamma _{Z^0}]},\\
P_{\gamma b}(s_{13})&=&P_{Z^0b}(s_{13}), \nonumber\\
P_{\gamma e}(s)&=&\frac{1}{s},\nonumber\\
P_{\gamma b}(s_{12})&=&P_{Z^0b}(s_{12}), \nonumber
\end{eqnarray}

\noindent and the corresponding tensors

\begin{eqnarray}
T^{\mu}_{Z^0b}&=&\bar u(k_{2})\gamma ^{\mu}(a_{b}-b_{b}\gamma _{5})(k\llap{/}_{1}+k_{3})v(k\llap{/}_{3}),\nonumber\\
T_{Z^0e \mu}&=&\bar v(p_{2})\gamma _{\mu}(a_{e}-b_{e}\gamma _{5})u(p_{1}),\nonumber\\
T^{\mu}_{Z^0b}&=&\bar u(k_{2})(k\llap{/}_{1}+k\llap{/}_{2})\gamma^{\mu}(a_{b}-b_{b}\gamma _{5})v(k_{3}),\nonumber\\
T^{\mu}_{A^0b}&=&\bar u(k_{2})\gamma _{5}(k_{2}+k_{3}-k_{1})^{\mu}v(k_{3}),\\
T^{\mu}_{Z^0b}&=&\bar u(k_{2})\gamma ^{\mu}(a_{b}-b_{b}\gamma _{5})v{k_{3}},\nonumber\\
T^{\mu}_{\gamma b}&=&\bar u(k_{2})\gamma ^{\mu}(k\llap{/}_{1}+k\llap{/}_{3})v(k_{3}),\nonumber\\
T_{\gamma b \mu}&=&\bar v(p_{2})\gamma _{\mu}u(p_{1}),\nonumber\\
T^{\mu}_{\gamma b}&=&\bar u(k_{2})(k\llap{/}_{1}+k\llap{/}_{2})\gamma ^{\mu}v(k_{3}).\nonumber
\end{eqnarray}

In fact, we rearrange the tensors $T^{'}$s in such a way that they become appropriate to
a computer program. Then, following the rules from helicity calculus formalism
\cite{11,12,13,14,15,16,17} and using identities of the type

\begin{equation}
\{\bar u_{\lambda}(p_{1})\gamma ^{\mu}u_{\lambda}(p_{2})\}\gamma_{\mu}=2u_{\lambda}(p_{2})\bar u_{\lambda}(p_{1})+2u_{-\lambda}(p_{1})\bar u_{-\lambda}(p_{2}),
\end{equation}

\noindent which is in fact the so called Chisholm identity, and

\begin{equation}
p\llap{/}=u_{\lambda}(p)\bar u_{\lambda}(p)+u_{-\lambda}(p)\bar u_{-\lambda}(p),
\end{equation}

\noindent defined as a sum of the two projections $u_{\lambda}(p)\bar u_{\lambda}(p)$
and $u_{-\lambda}(p)\bar u_{-\lambda}(p)$.

The spinor products are given by 

\begin{eqnarray}
s(p_{i}, p_{j})&\equiv&\bar u_{+}(p_{i})u_{-}(p_{j})=-s(p_{j}, p_{i}),\nonumber\\
t(p_{i}, p_{j})&\equiv&\bar u_{-}(p_{i})u_{+}(p_{j})=[s(p_{j}, p_{i})]^{*}.
\end{eqnarray}

Using Eqs. (9), (10) and (11), which are proved in Ref. \cite{17}, we
can reduce many amplitudes to expressions involving only spinor products.

Evaluating  the tensor products of (5) for each combination of $(\lambda, \lambda ^{'})$
with $\lambda, \lambda ^{'}= \pm 1$ one obtains the following expressions

\begin{eqnarray}
{\cal M}_{1}(+,+)&=&F_{1}f_{1}^{+,+}s(k_{2},p_{2})t(p_{1}, k_{1})s(k_{1}, k_{3}),\nonumber\\
{\cal M}_{1}(+,-)&=&F_{1}f_{1}^{+,-}s(k_{2},p_{1})t(p_{2}, k_{1})s(k_{1}, k_{3}),\nonumber\\
{\cal M}_{1}(-,+)&=&F_{1}f_{1}^{-,+}t(k_{2},p_{1})s(p_{2},k_{1})t(k_{1},k_{3}),\\
{\cal M}_{1}(-,-)&=&F_{1}f_{1}^{-,-}t(k_{2},p_{2})s(p_{1}, k_{1})t(k_{1},k_{3}),\nonumber \end{eqnarray}

\begin{eqnarray}
{\cal M}_{2}(+,+)&=&F_{2}f_{2}^{+,+}t(k_{2},k_{1})s(k_{1}, p_{2})t(p_{1}, k_{3}),\nonumber\\
{\cal M}_{2}(+,-)&=&F_{2}f_{2}^{+,-}t(k_{2},k_{1})s(k_{1}, p_{1})t(p_{2}, k_{3}),\nonumber\\
{\cal M}_{2}(-,+)&=&F_{2}f_{2}^{-,+}s(k_{2},k_{1})t(k_{1}, p_{1})s(p_{2}, k_{3}),\\
{\cal M}_{2}(-,-)&=&F_{2}f_{2}^{-,-}s(k_{2},k_{1})t(k_{1}, p_{2})s(p_{1}, k_{3}),\nonumber
\end{eqnarray}

\begin{eqnarray}
{\cal M}_{3}(+,+)&=&F_{3}f_{3}^{+,+}t(k_{2},k_{3})[ s(p_{2}, k_{2})t(k_{2},p_{1})+ s(p_{2}, k_{3})t(k_{3},p_{1})-s(p_{2},k_{1})t(k_{1},p_{1})],\nonumber\\
{\cal M}_{3}(+,-)&=&F_{3}f_{3}^{+,-}t(k_{2},k_{3})[ t(p_{2}, k_{2})s(k_{2},p_{1})+ t(p_{2}, k_{3})s(k_{3},p_{1})-t(p_{2},k_{1})s(k_{1},p_{1})],\nonumber\\
{\cal M}_{3}(-,+)&=&F_{3}f_{3}^{-,+}s(k_{2},k_{3})[ s(p_{2}, k_{2})t(k_{2},p_{1})+ s(p_{2}, k_{3})t(k_{3},p_{1})-s(p_{2},k_{1})t(k_{1},p_{1})],\\
{\cal M}_{3}(-,-)&=&F_{3}f_{3}^{-,-}s(k_{2},k_{3})[ t(p_{2}, k_{2})s(k_{2},p_{1})+ t(p_{2}, k_{3})s(k_{3},p_{1})-t(p_{2},k_{1})s(k_{1},p_{1})],\nonumber
\end{eqnarray}

\begin{eqnarray}
{\cal M}_{4}(+,+)&=&F_{4}f_{4}^{+,+}s(k_{2},p_{2})t(p_{1}, k_{3}),\nonumber\\
{\cal M}_{4}(+,-)&=&F_{4}f_{4}^{+,-}s(k_{2},p_{1})t(p_{2}, k_{3}),\nonumber\\
{\cal M}_{4}(-,+)&=&F_{4}f_{4}^{-,+}t(k_{2},p_{1})s(p_{2}, k_{3}),\\
{\cal M}_{4}(-,-)&=&F_{4}f_{4}^{-,-}t(k_{2},p_{2})s(p_{1}, k_{3}),\nonumber
\end{eqnarray}

\begin{eqnarray}
{\cal M}_{5}(+,+)&=&F_{5}s(k_{2},k_{1})t(k_{1}, p_{1})s(p_{2}, k_{3}),\nonumber\\
{\cal M}_{5}(+,-)&=&F_{5}s(k_{2},k_{1})t(k_{1}, p_{2})s(p_{1}, k_{3}),\nonumber\\
{\cal M}_{5}(-,+)&=&F_{5}t(k_{2},k_{1})s(k_{1}, p_{2})t(p_{1}, k_{3}),\\
{\cal M}_{5}(-,-)&=&F_{5}t(k_{2},k_{1})s(k_{1}, p_{1})t(p_{2}, k_{3}),\nonumber
\end{eqnarray}

\begin{eqnarray}
{\cal M}_{6}(+,+)&=&F_{6}s(k_{2},p_{2})t(p_{1}, k_{1})s(k_{1}, k_{3}),\nonumber\\
{\cal M}_{6}(+,-)&=&F_{6}s(k_{2},p_{1})t(p_{2}, k_{1})s(k_{1}, k_{3}),\nonumber\\
{\cal M}_{6}(-,+)&=&F_{6}t(k_{2},p_{1})s(p_{2}, k_{1})t(k_{1}, k_{3}),\\
{\cal M}_{6}(-,-)&=&F_{6}t(k_{2},p_{2})s(p_{1}, k_{1})t(k_{1}, k_{3}),\nonumber
\end{eqnarray}

\noindent where

\begin{eqnarray}
F_{1}&=&-2iC_{1}P_{Z^0b}(s_{13})P_{Z^0e}(s),\nonumber\\
F_{2}&=&-2iC_{2}P_{Z^0b}(s_{12})P_{Z^0e}(s),\nonumber\\
F_{3}&=&-iC_{3}P_{hb}(s_{23})P_{Z^0e}(s),\nonumber\\
F_{4}&=&2iC_{4}P_{Z^0b}(s_{23})P_{Z^0e}(s),\\
F_{5}&=&-2iC_{5}P_{\gamma b}(s_{13})P_{\gamma e}(s),\nonumber\\
F_{6}&=&-2iC_{6}P_{\gamma b}(s_{12})P_{\gamma e}(s),\nonumber
\end{eqnarray}

\noindent and

\begin{eqnarray}
f_{1}^{+,+}&=&f_{2}^{+,+}=f_{4}^{+,+}=(a_{b}-b_{b})(a_{e}-b_{e}),\nonumber\\
f_{1}^{+,-}&=&f_{2}^{+,-}=f_{4}^{+,-}=(a_{b}-b_{b})(a_{e}+b_{e}),\nonumber\\
f_{1}^{-,+}&=&f_{2}^{-,+,}=f_{4}^{-,+}=(a_{b}+b_{b})(a_{e}-b_{e}),\nonumber\\
f_{1}^{-,-}&=&f_{2}^{-,-}=f_{4}^{-,-}=(a_{b}+b_{b})(a_{e}+b_{e}),\\
f_{3}^{+,+}&=&f_{3}^{-,+,}=(a_{e}-b_{e}),\nonumber\\
f_{3}^{+,-}&=&f_{3}^{-,-}=(a_{e}+b_{e}).\nonumber
\end{eqnarray}

\noindent Here, $a_{e}=-1+4\sin ^{2}\theta _{W}$, $b_{e}=1$, $a_{b}=-1+\frac{4}{3}\sin ^{2}\theta_{W}$
and $b_{b}=-1$, according to the experimental data \cite{22}.

\subsection{Case $b\bar b A^0$}

The Feynman diagrams that contribute to the process $e^{+}e^{-}
\rightarrow b\bar b A^0$ to tree-level, are shown in Fig. 2. The
corresponding amplitudes to each graph are:

\begin{eqnarray}
{\cal M}_1&=&C_{1}P_{Z^0b}(s_{13})P_{Z^0e}(s)T_{Z^0b}^{\mu}T_{Z^0e \mu},\nonumber\\
{\cal M}_2&=&C_{2}P_{Z^0b}(s_{12})P_{Z^0e}(s)T_{Z^0b}^{\mu}T_{Z^0e \mu},\nonumber\\
{\cal M}_3&=&C_{3}P_{Z^0H}(s_{23})P_{Z^0e}(s)T^{\mu}_{Z^0H}T_{Z^0e \mu},\nonumber\\
{\cal M}_4&=&C_{4}P_{Z^0h}(s_{23})P_{Z^0e}(s)T^{\mu}_{Z^0h}T_{Z^0e \mu},\\
{\cal M}_5&=&C_{5}P_{\gamma b}(s_{13})P_{\gamma e}(s)T^{\mu}_{\gamma b}T_{\gamma e \mu},\nonumber\\
{\cal M}_6&=&C_{6}P_{\gamma b}(s_{12})P_{\gamma e}(s)T^{\mu}_{\gamma b}T_{\gamma e \mu},\nonumber
\end{eqnarray}

\noindent where

\begin{eqnarray}
C_{1}&=&g^{3}m_{b}\tan \beta /32M_{W}\cos ^{2}\theta_{W},\nonumber\\
C_{2}&=&C_{1},\nonumber\\
C_{3}&=&-g^{3}m_{b}\cos \alpha \sin (\beta-\alpha) /16M_{W}\cos \beta \cos ^{2}\theta_{W},\nonumber\\
C_{4}&=&-g^{3}m_{b}\sin \alpha \cos (\beta-\alpha) /16M_{W}\cos \beta \cos ^{2}\theta_{W},\\
C_{5}&=&gm_{b}e^{2}Q_{e}Q_{b}\tan \beta /2M_{W},\nonumber\\
C_{6}&=&C_{5},\nonumber
\end{eqnarray}
The propagators are given in Eq. (7), while the tensors $T^{\mu}_{Z^0b}$ and
$T_{Z^0e \mu}$ Eq. (8), and the new ones as follows.

\begin{eqnarray}
T^{\mu}_{Z^0b}&=&\bar u(k_{2})\gamma_{5}(k\llap{/}_{1}+k\llap{/}_{2})\gamma^{\mu}(a_{b}-b_{b}\gamma _{5})v(k_{3}),\nonumber\\
T^{\mu}_{Z^0h}&=&\bar u(k_{2})(k_{1}-k_{2}-k_{3})^{\mu}v(k_{3}).
\end{eqnarray}

Making use of Eqs. (9)-(11), we can reduce the amplitude ${\cal M}$ to expressions that
contain only spinor products. Finally, evaluating the tensor products in Eq. (20)
for each combination of $\lambda , \lambda ^{'} = \pm 1$ we obtain:

\begin{eqnarray}
{\cal M}_{1}(+,+)&=&2G_{1}g_{1}^{+,+}t(k_{2},k_{1})s(k_{1},p_{2})t(p_{1},k_{3}),\nonumber\\
{\cal M}_{1}(+,-)&=&2G_{1}g_{1}^{+,-}t(k_{2},k_{1})s(k_{1},p_{1})t(p_{2},k_{3}),\nonumber\\
{\cal M}_{1}(-,+)&=&2G_{1}g_{1}^{-,+}s(k_{2},k_{1})t(k_{1},p_{1})s(p_{2},k_{3}),\\
{\cal M}_{1}(-,-)&=&2G_{1}g_{1}^{-,-}s(k_{2},k_{1})t(k_{1},p_{2})s(p_{1},k_{3}),\nonumber
\end{eqnarray}

\begin{eqnarray}
{\cal M}_{2}(+,+)&=&2G_{2}g_{2}^{+,+}t(k_{2},p_{1})s(p_{2},k_{1})t(k_{1},k_{3}),\nonumber\\
{\cal M}_{2}(+,-)&=&2G_{2}g_{2}^{+,-}t(k_{2},p_{2})s(p_{1},k_{1})t(k_{1},k_{3}),\nonumber\\
{\cal M}_{2}(-,+)&=&2G_{2}g_{2}^{-,+}s(k_{2},p_{2})t(p_{1},k_{1})s(k_{1},k_{3}),\\
{\cal M}_{2}(-,-)&=&2G_{2}g_{2}^{-,-}s(k_{2},p_{1})t(p_{2},k_{1})s(k_{1},k_{3}),\nonumber
\end{eqnarray}

\begin{eqnarray}
{\cal M}_{3}(+,+)&=&-G_{3}g_{3}^{+,+}s(k_{2},k_{3})[t(p_{2},k_{2})s(k_{2},p_{1})+t(p_{2},k_{3})s(k_{3},p_{1})-t(p_{2},k_{1})s(k_{1},p_{1})],\nonumber\\
{\cal M}_{3}(+,-)&=&-G_{3}g_{3}^{+,-}s(k_{2},k_{3})[s(p_{2},k_{2})t(k_{2},p_{1})+s(p_{2},k_{3})t(k_{3},p_{1})-s(p_{2},k_{1})t(k_{1},p_{1})],\nonumber\\
{\cal M}_{3}(-,+)&=&-G_{3}g_{3}^{-,+}t(k_{2},k_{3})[t(p_{2},k_{2})s(k_{2},p_{1})+t(p_{2},k_{3})s(k_{3},p_{1})-t(p_{2},k_{1})s(k_{1},p_{1})],\\
{\cal M}_{3}(-,-)&=&-G_{3}g_{3}^{-,-}t(k_{2},k_{3})[s(p_{2},k_{2})t(k_{2},p_{1})+s(p_{2},k_{3})t(k_{3},p_{1})-s(p_{2},k_{1})t(k_{1},p_{1})],\nonumber
\end{eqnarray}

\begin{eqnarray}
{\cal M}_{4}(+,+)&=&-G_{4}g_{4}^{+,+}t(k_{2},k_{3})[s(p_{2},k_{2})t(k_{2},p_{1})+s(p_{2},k_{3})t(k_{3},p_{1})-s(p_{2},k_{1})t(k_{1},p_{1})],\nonumber\\
{\cal M}_{4}(+,-)&=&-G_{4}g_{4}^{+,-}t(k_{2},k_{3})[t(p_{2},k_{2})s(k_{2},p_{1})+t(p_{2},k_{3})s(k_{3},p_{1})-t(p_{2},k_{1})s(k_{1},p_{1})],\nonumber\\
{\cal M}_{4}(-,+)&=&-G_{4}g_{4}^{-,+}s(k_{2},k_{3})[s(p_{2},k_{2})t(k_{2},p_{1})+s(p_{2},k_{3})t(k_{3},p_{1})-s(p_{2},k_{1})t(k_{1},p_{1})],\\
{\cal M}_{4}(-,-)&=&-G_{4}g_{4}^{-,-}s(k_{2},k_{3})[t(p_{2},k_{2})s(k_{2},p_{1})+t(p_{2},k_{3})s(k_{3},p_{1})-t(p_{2},k_{1})s(k_{1},p_{1})],\nonumber
\end{eqnarray}

\begin{eqnarray}
{\cal M}_{5}(+,+)&=&-2G_{5}t(k_{2},p_{1})s(p_{2},k_{1})t(k_{1},k_{3}),\nonumber\\
{\cal M}_{5}(+,-)&=&-2G_{5}t(k_{2},p_{2})s(p_{1},k_{1})t(k_{1},k_{3}),\nonumber\\
{\cal M}_{5}(-,+)&=&2G_{5}s(k_{2},p_{2})t(p_{1},k_{1})s(k_{1},k_{3}),\\
{\cal M}_{5}(-,-)&=&2G_{5}s(k_{2},p_{1})t(p_{2},k_{1})s(k_{1},k_{3}),\nonumber
\end{eqnarray}

\begin{eqnarray}
{\cal M}_{6}(+,+)&=&2G_{6}t(k_{2},k_{1})s(k_{1},p_{2})t(p_{1},k_{3}),\nonumber\\
{\cal M}_{6}(+,-)&=&2G_{6}t(k_{2},k_{1})s(k_{1},p_{1})t(p_{2},k_{3}),\nonumber\\
{\cal M}_{6}(-,+)&=&-2G_{6}s(k_{2},k_{1})t(k_{1},p_{1})s(p_{2},k_{3}),\\
{\cal M}_{6}(-,-)&=&-2G_{6}s(k_{2},k_{1})t(k_{1},p_{2})s(p_{1},k_{3}),\nonumber
\end{eqnarray}

\noindent where

\begin{eqnarray}
G_{1}&=&C_{1}P_{Z^0b}(s_{13})P_{Z^0e}(s),\nonumber\\
G_{2}&=&C_{2}P_{Z^0b}(s_{12})P_{Z^0e}(s),\nonumber\\
G_{3}&=&C_{3}P_{Z^0H}(s_{23})P_{Z^0e}(s),\nonumber\\
G_{4}&=&C_{4}P_{Z^0h}(s_{23})P_{Z^0e}(s),\\
G_{5}&=&C_{5}P_{\gamma b}(s_{13})P_{\gamma e}(s),\nonumber\\
G_{6}&=&C_{6}P_{\gamma b}(s_{12})P_{\gamma e}(s),\nonumber
\end{eqnarray}

\noindent and

\begin{eqnarray}
g_{1}^{+,+}&=&g_{2}^{+,+}= f_{1}^{+,+},  \mbox{\hspace{6mm}   $g_{3}^{+,+}=g_{4}^{+,+}=f_{3}^{+,+,}$},\nonumber\\
g_{1}^{+,-}&=&g_{2}^{+,-}= f_{1}^{+,-},  \mbox{\hspace{6mm}   $g_{3}^{+,-}=g_{4}^{+,-}=f_{3}^{+,-}$},\nonumber\\
g_{1}^{-,+}&=&g_{2}^{-,+}= f_{1}^{-,+},  \mbox{\hspace{6mm}   $g_{3}^{-,+}=g_{4}^{-,+}=f_{3}^{-,+}$},\\
g_{1}^{-,-}&=&g_{2}^{-,-}= f_{1}^{-,-},  \mbox{\hspace{6mm}   $g_{3}^{-,-}=g_{4}^{-,-}=f_{3}^{-,-}$}.\nonumber
\end{eqnarray}

\noindent The expressions for the $f_{1}^{+,+}, f_{1}^{+,-}, f_{1}^{-,+},
f_{1}^{-,-}, f_{3}^{+,+}, f_{3}^{+,-}, f_{3}^{-,+}$ and $f_{3}^{-,-}$ are
given in Eq. (19).

After the evaluation of the amplitudes of the corresponding diagrams, we obtain
the cross sections of the analyzed processes for each point of the phase
space using Eqs. (12)-(17) and (23)-(28) by a computer program, which makes
use of the subroutine RAMBO (Random Momenta Beautifully Organized). The
advantages of this procedure in comparison to the traditional "trace
technique" are discussed in Ref. \cite{11,12,13,14,15,16,17}.

We use the Breit-Wigner propagators for the $Z^0$, $h^0$, $H^0$ and $A^0$
bosons. The mass of the bottom $(m_{b}\approx 4.5\hspace*{2mm} GeV)$, the mass $(M_{Z^0} = 91.2 \hspace*{2mm}GeV)$
and width $(\Gamma_{Z^0} = 2.4974 \hspace*{2mm}GeV)$ of $Z^0$ have been taken as inputs;
the widths of $h^0$, $H^0$ and $A^0$ are calculated from the formulae
given in Ref. \cite{6}. In the next sections we present the numerical
computation of the processes $e^{+}e^{-}\rightarrow b \bar b h$,
$h=h^0,H^0,A^0$.

\section{Detection of Neutral Higgs Bosons at LEP-II and NLC Energies}

In an earlier paper \cite{21} has been explored the possibility of  finding one or more
of the neutral Higgs bosons predicted by the MSSM in $gg\rightarrow b\bar
b h$ $(h=h^0, H^0, A^0)$ followed by $h\rightarrow b\bar b$,
profiting from the very high $b$-tagging efficiencies. In other works \cite{18},
it has been studied the discovery reach of the Tevatron and the LHC for
detecting a Higgs boson $(h)$ via the processes $p\bar p/pp \rightarrow
b\bar bh (\rightarrow b \bar b)+X$ and it has been shown the possibility of
detecting SUSY Higgs bosons at FNAL and LHC if $\tan \beta$ is large.

In this paper, we study the detection of neutral MSSM Higgs bosons at
$e^{+}e^{-}$ colliders, including 3-body mode diagrams (Figs. 1.1, 1.2, 1.3,
1.5 and 1.6; Figs. 2.1, 2.2, 2.3, 2.5 and 2.6) besides the dominant mode
diagram (Fig. 1.4; Fig. 2.4) assuming an
integrated luminosity of ${\cal L} = 500$ $pb^{-1}$ and ${\cal L} = 10$ $fb^{-1}$
at $\sqrt{s}= 200$ $GeV$ and $500$ $GeV$ for LEP-II and NLC, respectively. We consider the
complete set of Feynman diagrams (Fig. 1-2) at tree-level and utilize the helicity formalism for
the evaluation of their amplitudes. In the next subsections, we present our results for the
case of the different Higgs bosons.

\subsection{Detection of $h^0$}

In order to illustrate our results on the detection of the $h^0$ Higgs boson, 
we present graphs in the parameters space region $(m_{A^0}-\tan\beta)$, assuming 
$m_{t}= 175$ $GeV$, $M_{\stackrel{\sim}t}= 500$ $GeV$ and $\tan\beta > 1$ for
LEP-II and NLC. Our results are
displayed in Fig. 3, for $e^{+}e^{-}\rightarrow (A^0, Z^0) + h^0$ dominant mode and
for the processes at 3-body $e^{+}e^{-}\rightarrow b\bar b h^0$.

The total cross section for each contour is $0.01$ $pb$, $0.02$ $pb$ and $0.05$ $pb$, which gives 5
events, 10 events and 25 events, respectively. We can see from this figure, that the effect of the 
reaction $b\bar b h^0$ is lightly more important that $(A^0, Z^0) + h^0$, for most
of the $(m_{A^0}-\tan\beta)$ parameter space regions. Nevertheless, there are substantial
portions of parameters space in which the discovery of the $h^0$ is not possible using
either $(A^0, Z^0) + h^0$ or $b\bar b h^0$.

For the case of NLC, the results on the detection of the $h^0$ are shown in Fig. 4. It is
clear from this figure that the contribution of the process $e^{+}e^{-}\rightarrow b\bar b h^0$
becomes dominant, namely $e^{+}e^{-}\rightarrow (A^0, Z^0) + h^0$ is small 
in all parameter space. However, they could provide important information on the
Higgs bosons detection. For instance, we give the contours for the total
cross section to say
0.01 $pb$ and 0.03 $pb$ for both processes. These cross sections give 100 events and 300
events in total to each process, then it will be detectable the $h^0$ at NLC energies. 

\subsection{Detection of $H^0$}

The detection of the heavy Higgs bosons $H^0$ is not possible at LEP-II. Nevertheless,
the possibility for a future $e^{+}e^{-}$ collider with center of mass energy of order 500 $GeV$
is surveyed. Consequently, detection of the $H^0$ and $A^0$ will be only possible
in such a scenario if $\sqrt{s}$ is significantly larger than $m_{A^0}+m_{H^0}$, {\it i.e.}
$>\!\!\!\!\!\! _{\sim} 2m_{A^0}$ for large $m_{A^0}$ where $m_{H^0}\approx m_{A^0}$.
To illustrate more precisely our results, we give the contours for the total cross section for 
both processes $e^{+}e^{-}\rightarrow (A^0, Z^0) + H^0$, $e^{+}e^{-}\rightarrow b\bar b H^0$ in Fig. 5 
for an NLC with $\sqrt{s}= 500$ $GeV$ and ${\cal L} = 10$ $fb^{-1}$ in the case where $m_{t}=175$ $GeV$
and $M_{\stackrel{\sim}t}= 500$ $GeV$ (squark mixing is neglected). The contours 
for this cross section are 0.01 $pb$, 0.001 $pb$ and 0.0001 $pb$ for both reactions $(A^0, Z^0) + H^0$ and
$b\bar b H^0$. While that the number of events corresponding to each contour are 100,
10 and 1, respectively.

Our estimate is that if more than 100 total events are obtained for a given process
$(A^0, Z^0) + H^0 \hspace*{2mm} or \hspace*{2mm}b\bar b H^0$ then the Higgs boson $H^0$ can be detectable. Contours
for 10 events are also shown, but detection of any of the two cases with so few events
would require very high experimental and analysis efficiencies.

The effect of incorporate $b\bar b H^0$ in the detection of the Higgs boson $H^0$
is more important than the case of 2-body mode $(A^0, Z^0)+H^0$, because $b\bar b H^0$ cover
a major region in the parameters space $(m_{A^0}-\tan\beta)$. The most important
conclusion from this figure is that detection of all of the neutral Higgs bosons will be
possible at $\sqrt{s}=500$ $GeV$.

\subsection{Detection of $A^0$}

For the pseudoscalar $A^0$, it is interesting to consider the production mode
into $b\bar b A^0$, since it can have large cross section due to that the coupling of
$A^0$ with $b\bar b$ is directly proportional to $\tan\beta$, thus will always grow
with it. In Fig. 6, we present the contours of the cross sections for  the
process of our interest $b\bar b A^0$, at LEP-II energies.

We display the contour lines for $\sigma= 0.01, 0.02$, showing also the regions where the $A^0$
can be detected. These cross sections give a contour of production of 5 and 10 events. It is
clear from this figure that to detecting  the Higgs boson $A^0$ are necessary very 
high experimental and analysis efficiencies.

On the other hand, if we focus the detection of the $A^0$ at Next Linear $e^{+}e^{-}$ Collider
with $\sqrt{s} = 500$ $GeV$ and ${\cal L} = 10$ $fb^{-1}$, the panorama for its detection is 
more extensive. The Fig. 7, show the contours lines in the plane $(m_{A^0}-\tan\beta)$, to the
cross section of $b\bar bA^0$. The contours for this 
cross section correspond to 100, 30 and 10 events. It is estimated that if more than 100
total events are obtained for $b\bar b A^0$, then it is possible to detect the $A^0$.

\section{Conclusions}

In this paper, we have calculated the production of a neutral Higgs bosons in 
association with $b$-quarks via the processes $e^{+}e^{-}\rightarrow b \bar b h$,
$h = h^0, H^0, A^0$ and using the helicity formalism. We find that this 
processes could help to detect a possible neutral Higgs boson at LEP-II and
NLC energies when $\tan \beta$ is large.

The detection of $h^0$ through of the reaction $e^{+}e^{-}\rightarrow b \bar b h^0$,
compete favorable with the mode dominant $e^{+}e^{-}\rightarrow (A^0, Z^0) + h^0$. 
The process at 3-body $b\bar b h^0$ cover lightly a major portion of the parameter
space $(m_{A^0}-\tan \beta)$ as is shown in Fig. 3 and the corresponding
cross section for each contour is of $\sigma =$ 0.01 $pb$, 0.02 $pb$, 0.05 $pb$  for LEP-II; while 
for NLC we have that $\sigma =$ 0.01 $pb$, 0.03 $pb$  and the corresponding
contours are show in Fig. 4. However, there is a portion of the plane $(m_{A^0}-\tan\beta)$ where $h^0$
is not detectable  with the mechanism $(A^0, Z^0) + h^0$ or $b\bar b h^0$.

The heavy Higgs boson $H^0$ is not detectable at LEP-II energies. Nevertheless,
it could be detected at NLC via the $(A^0, Z^0) + H^0$ or $b\bar b h^0$ reaction,
with $H^0$ being produced in association with $b$-quarks. The detection of the $H^0$
will only be possible  in such scenarios if $\sqrt{s}$ is significantly
larger that $m_{A^0}+m_{H^0}$, ${\it i.e.}$ $>\!\!\!\!\!\! _{\sim} 2m_{A^0}$
for large $m_{A^0}$ where $m_{H^0}\approx m_{A^0}$. In Fig. 5, we give  
the contours for the cross sections for both cases $(A^0, Z^0) + H^0$ and $b\bar b H^0$.
We find that for $m_{A^0}$ and $\tan \beta$ large the reaction $e^{+}e^{-}\rightarrow b\bar b H^0$
is more important that $e^{+}e^{-}\rightarrow (A^0, Z^0)+H^0$ and cover
a major region in the parameter space $(m_{A^0}-\tan\beta)$ where this mode is
kinematically allowed.

It is interesting to notice that the associate production of $A^0$ with $b\bar b$ is
enhanced for larger values of $\tan \beta$, which could be used to detect neutral Higgs
bosons, provided that it will be possible also to tag the signals using the $b$-quarks
produced in the reaction. 

In the Figs. 6-7, we show the contours for the total cross section of the process
$b\bar b A^0$ for both LEP-II and NLC energies. We can conclude that there is a 
region where the Higgs boson $A^0$ could be detected at the next high energy
machines (NLC).

In summary, we conclude that the possibilities of detecting or excluding the 
neutral Higgs bosons of the Minimal Supersymmetric Standard Model $(h^0, H^0, A^0)$ in
the processes $e^{+}e^{-}\rightarrow b\bar b h$, $h = h^0, H^0, A^0$ are
important and in some cases are compared favorable with the mode dominant
$e^{+}e^{-}\rightarrow (A^0, Z^0) + h$, $h = h^0, H^0, A^0$ in the region of
parameters space $(m_{A^0}-\tan \beta)$ with $\tan \beta$ large. The detection of
the Higgs boson will require the combined use of the future high energy machine like 
LEP-II and the Next Linear $e^{+}e^{-}$ Collider.

\hspace{2cm}

\begin{center}
{\bf Acknowledgments}
\end{center}
This work was supported in part by {\it Consejo Nacional de Ciencia y
Tecnolog\'{\i}a} (CONACyT) and {\it Sistema Nacional de Investigadores}
(SNI) (M\'exico). O. A. S. would like to thank CONICET (Argentina).
We would like to thank J. L. D\'{\i}az-Cruz for suggestions and
for careful reading of our manuscript.

\newpage

\begin{center}
{\bf FIGURE CAPTIONS}
\end{center}

\vspace{5mm}

\bigskip

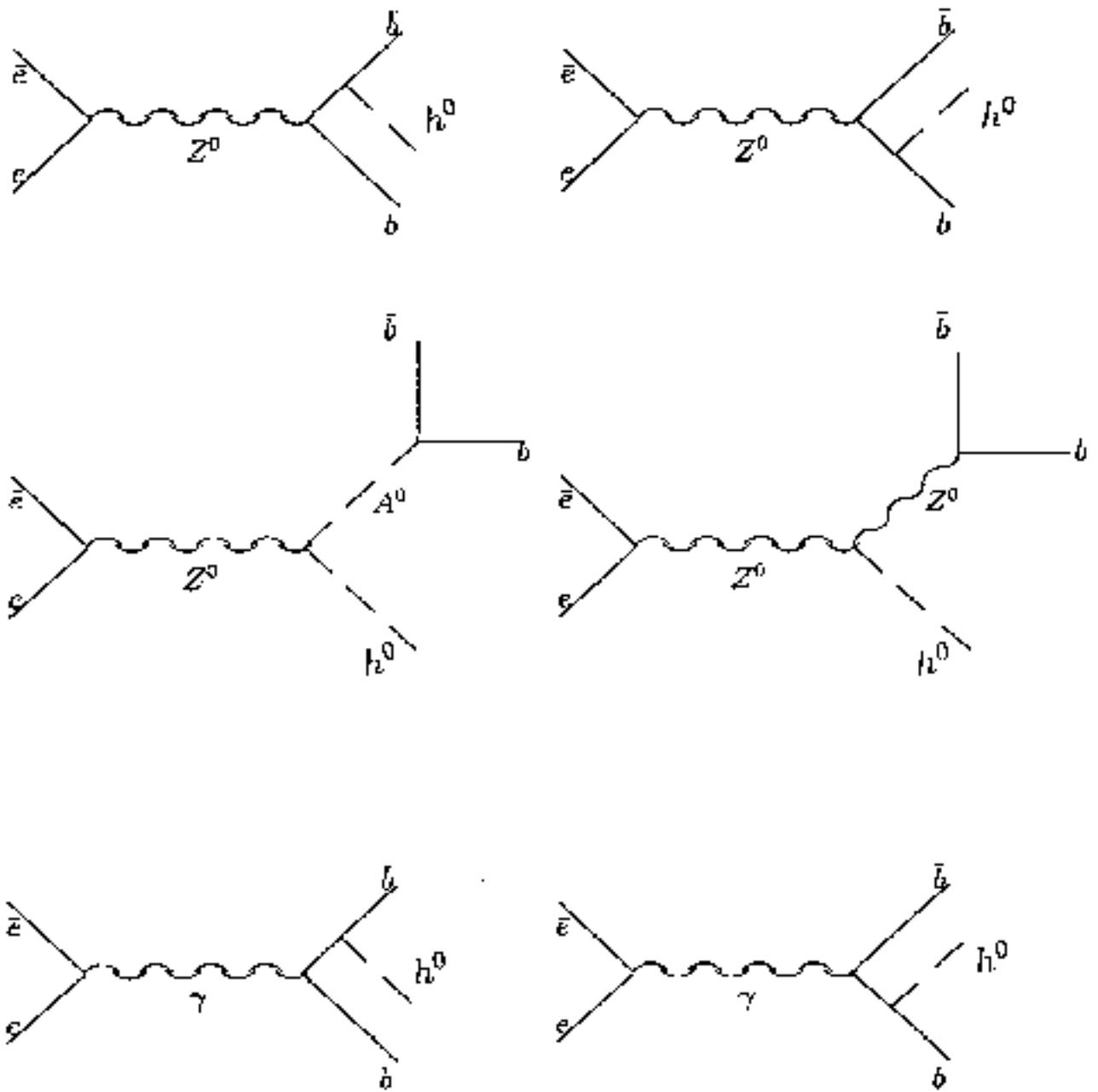
\noindent {\bf Fig. 1} Feynman Diagrams at tree-level for $e^{+}e^{-}
\rightarrow b\bar b h^0$. For $e^{+}e^{-}\rightarrow b\bar b H^0$
one has to make only the change $\sin\alpha / \cos\beta \rightarrow
\cos\alpha / \cos\beta$.

\bigskip

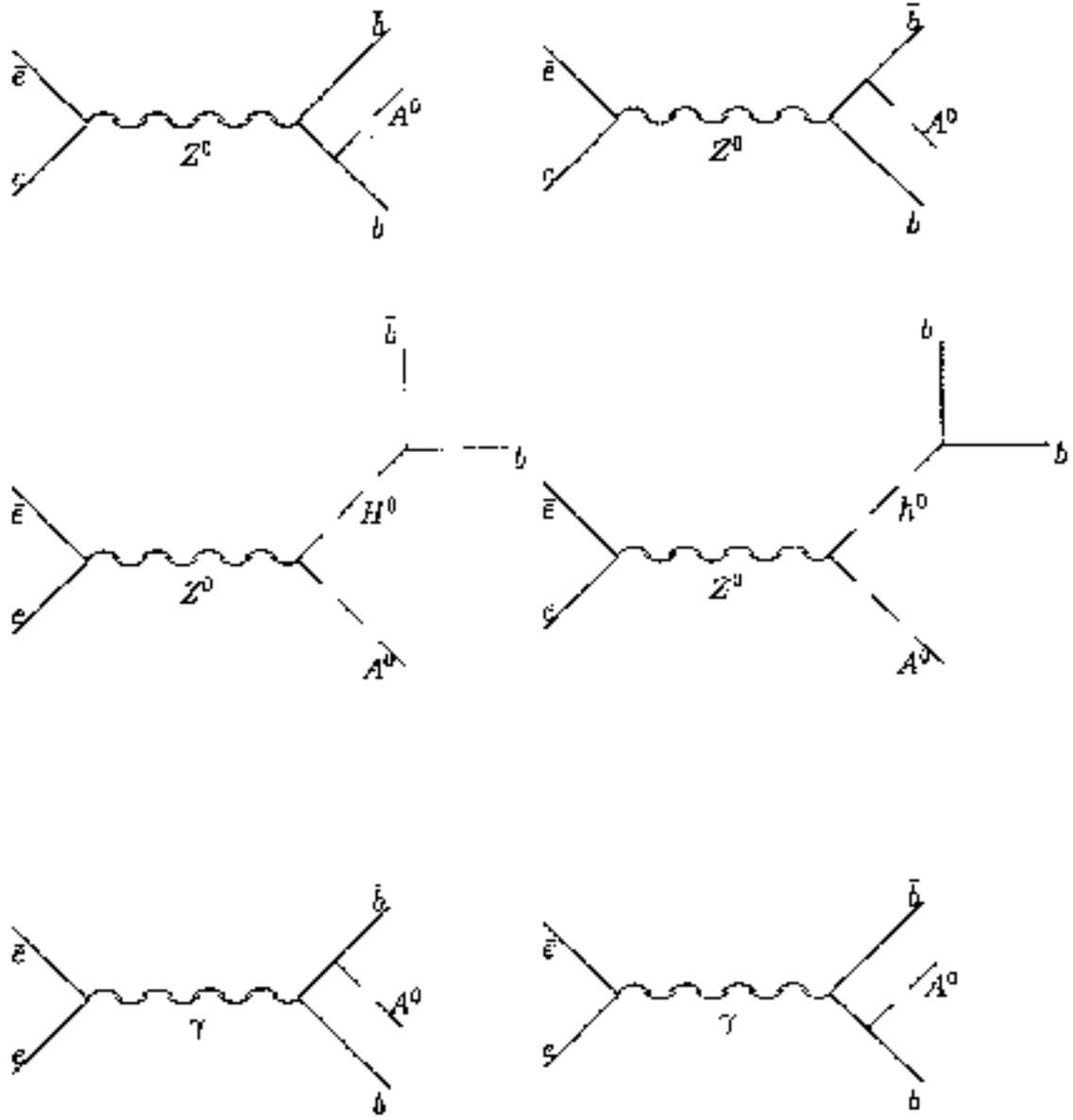
\noindent {\bf Fig. 2} Feynman Diagrams at tree-level for $e^{+}e^{-}
\rightarrow b\bar b A^0$.

\bigskip

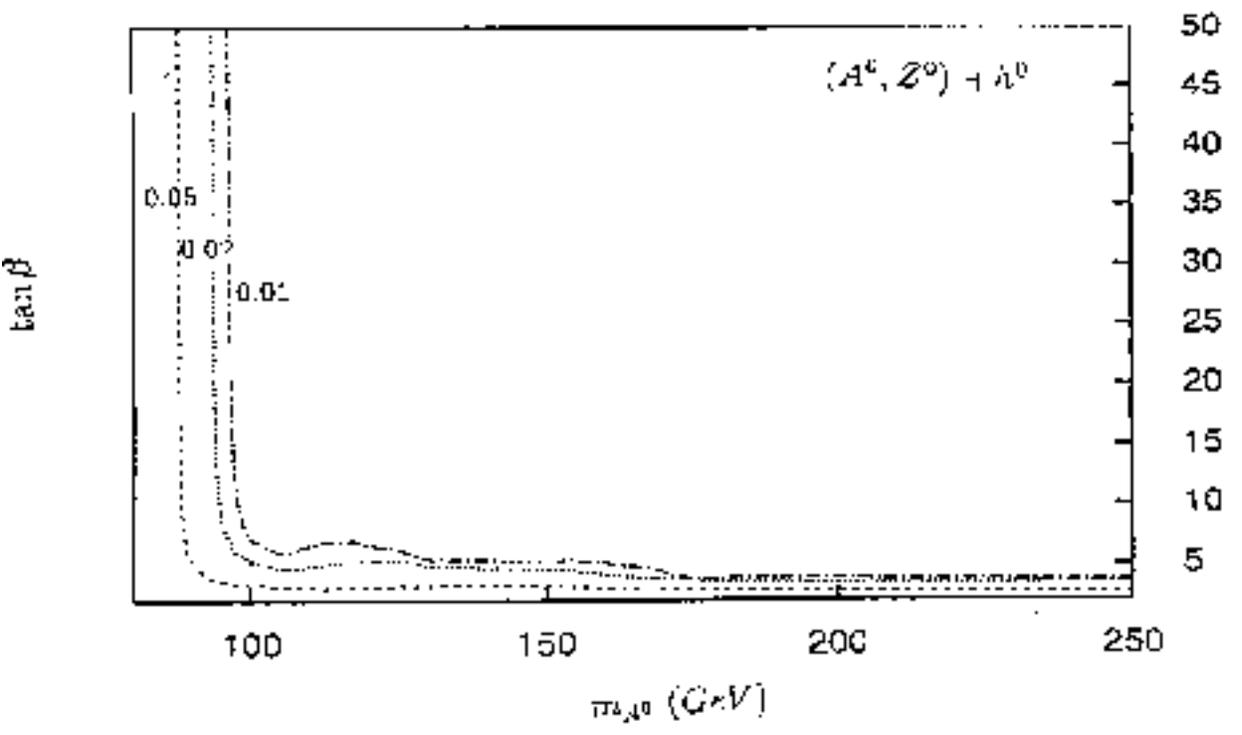
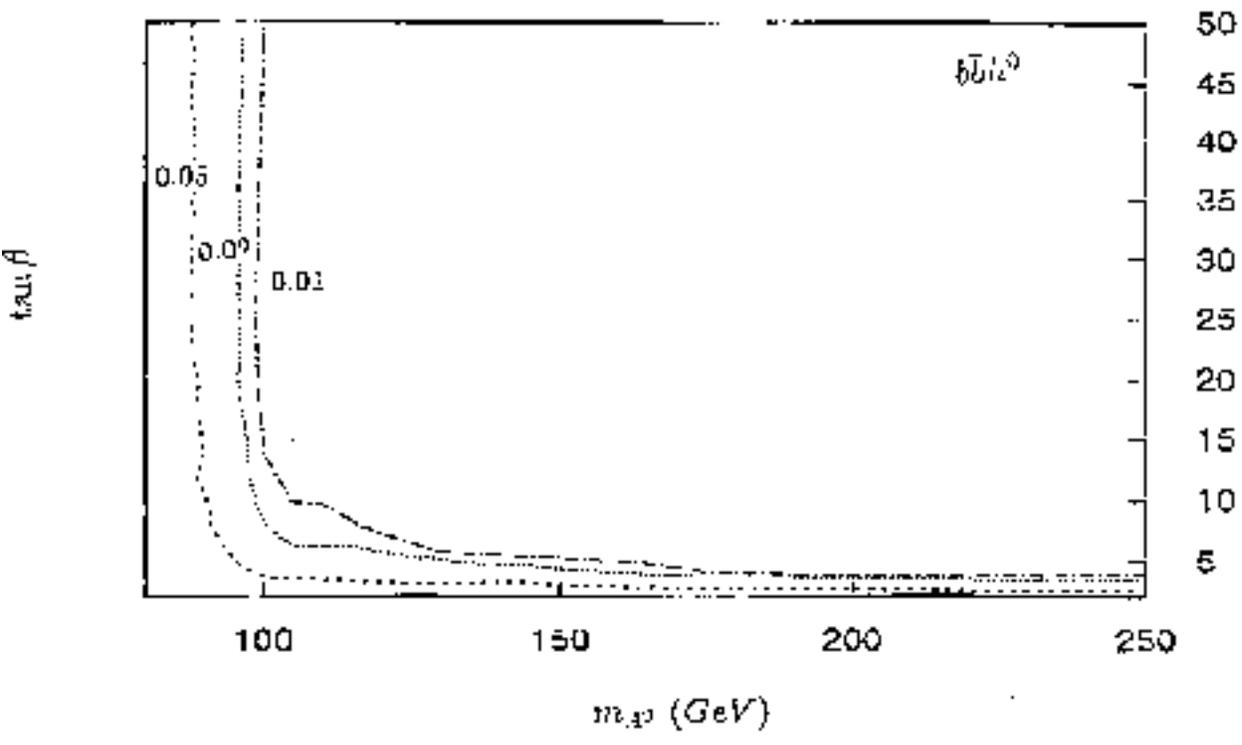
\noindent {\bf Fig. 3} Total cross sections contours in $(m_{A^0}-
\tan\beta)$ parameter space for $e^{+}e^{-}\rightarrow (A^0, Z^0) + h^0$
and $e^{+}e^{-}\rightarrow b\bar b h^0$ at LEP-II with $\sqrt{s} = 200$
$GeV$ and an integrated luminosity of ${{\cal L} = 500}$ $pb^{-1}$. We have
taken $m_{t} = 175$ $GeV$ and $M_{\stackrel {\sim}t} = 500$ $GeV$ and neglected
squark mixing.

\bigskip

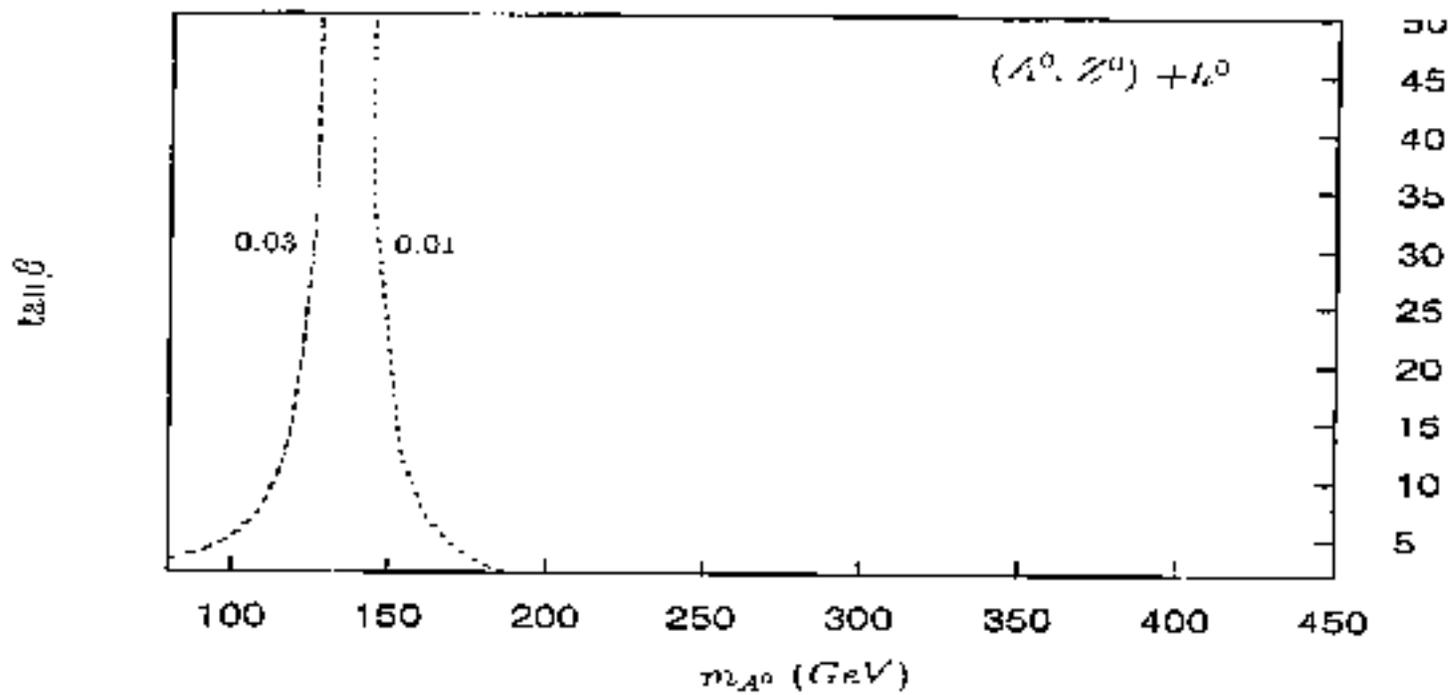
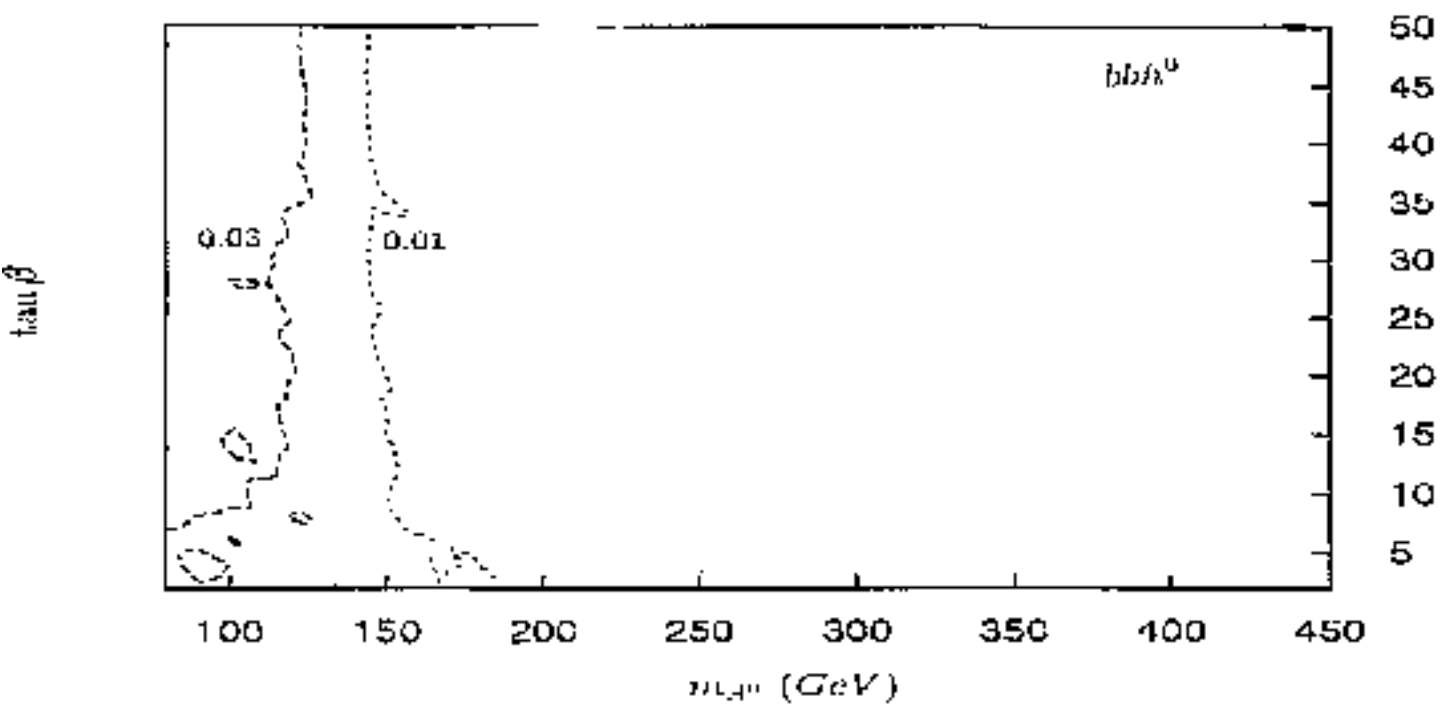
\noindent {\bf Fig. 4} Total cross sections contours for an NLC with
$\sqrt{s}= 500$ $GeV$ and ${\cal L} = 10$ $fb^{-1}$. We have taken
$m_{t} = 175$ $GeV$, $M_{\stackrel {\sim} t} = 500$ $GeV$ and neglected
squark mixing. We display contours for $e^{+}e^{-}\rightarrow (A^0, Z^0) +
h^0$ and $e^{+}e^{-}\rightarrow b\bar b h^0$, in the parameters space
$(m_{A^0}-\tan\beta)$.

\bigskip

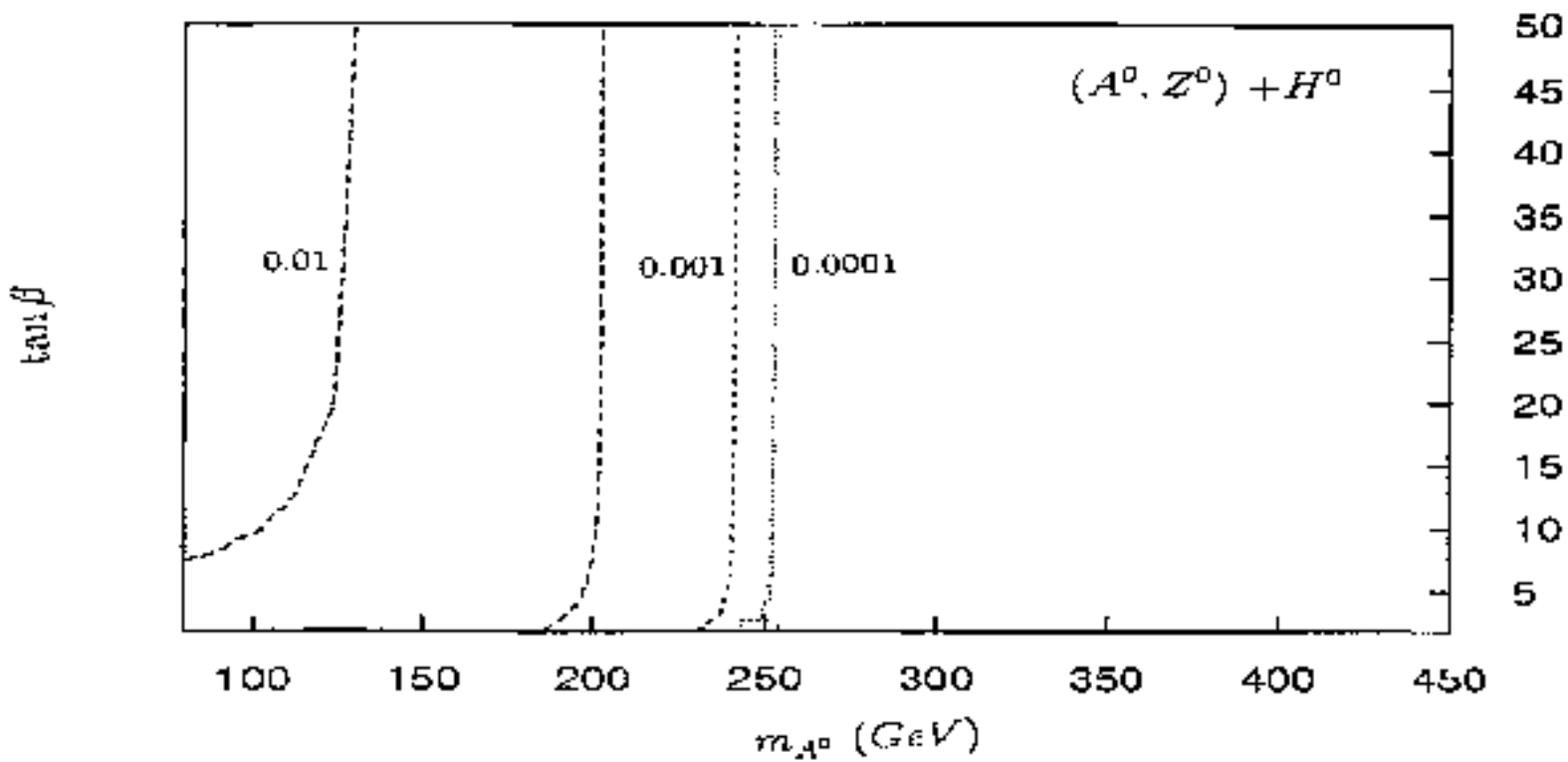
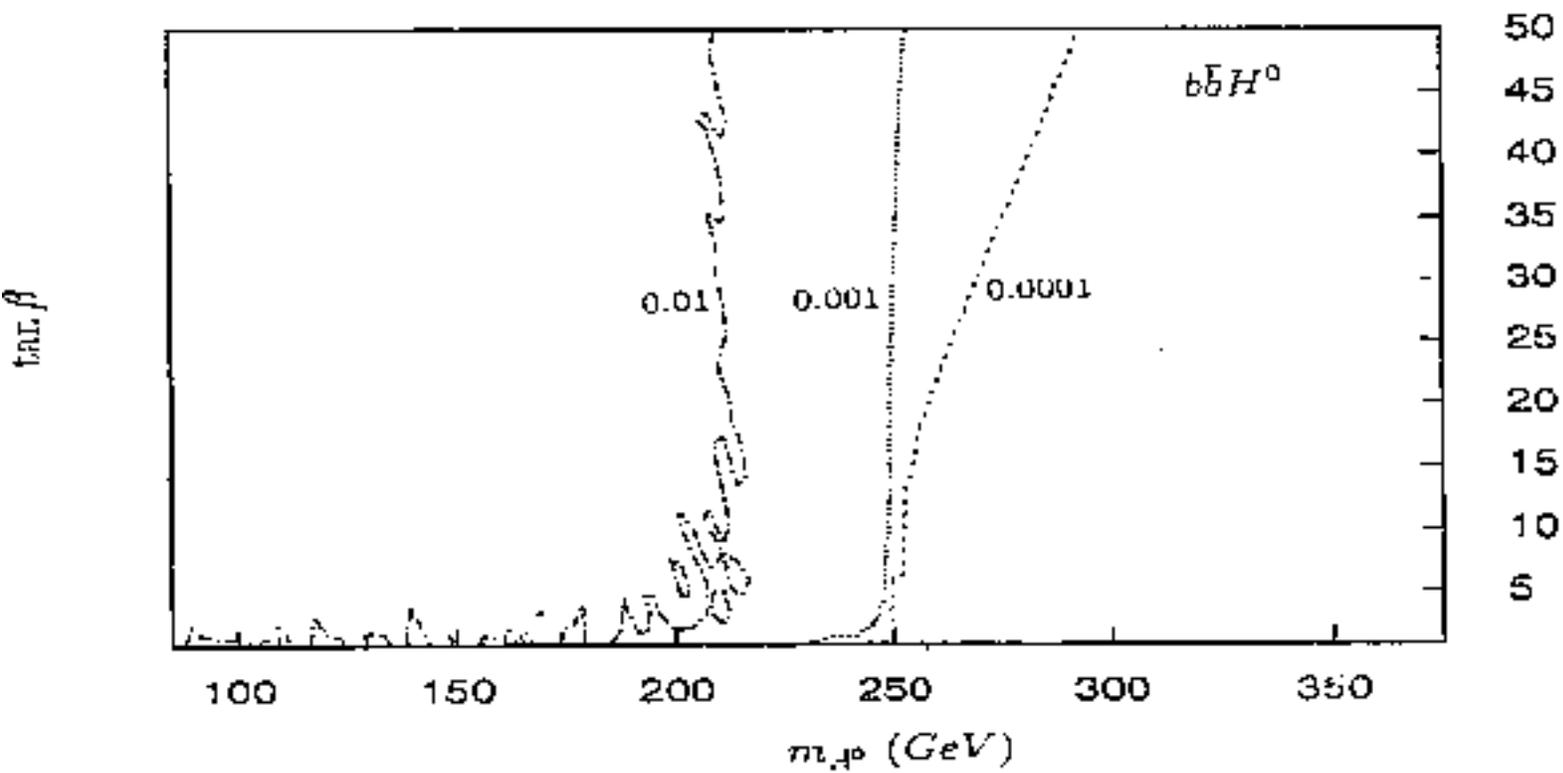
\noindent {\bf Fig. 5} Same as in Fig. 4, but for
$e^{+}e^{-}\rightarrow (A^0, Z^0) + H^0$ and $e^{+}e^{-}\rightarrow b
\bar b H^0$.

\bigskip

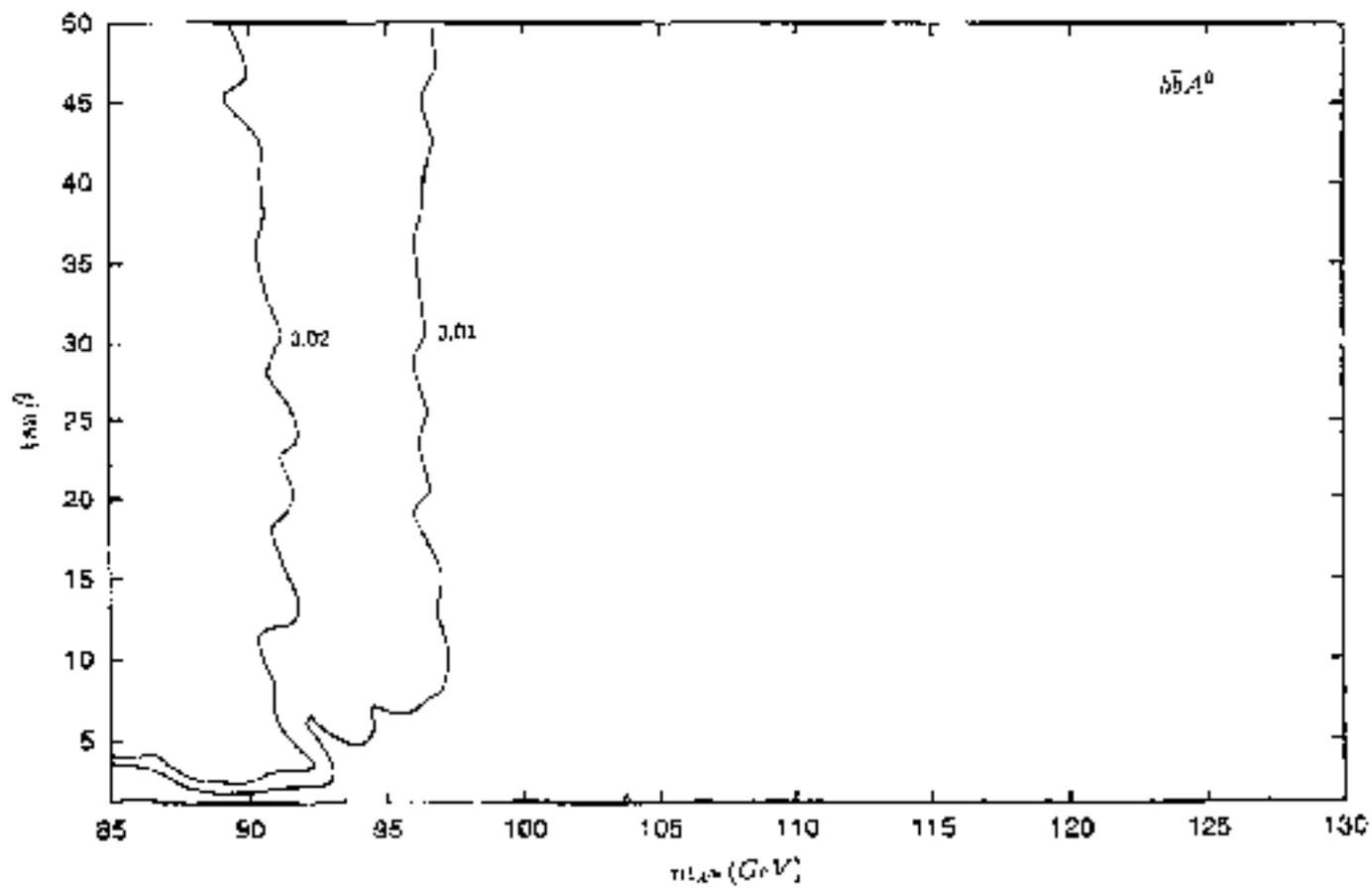
\noindent {\bf Fig. 6} Same as in Fig. 3, but for 
$e^{+}e^{-}\rightarrow b\bar b A^0$.

\bigskip

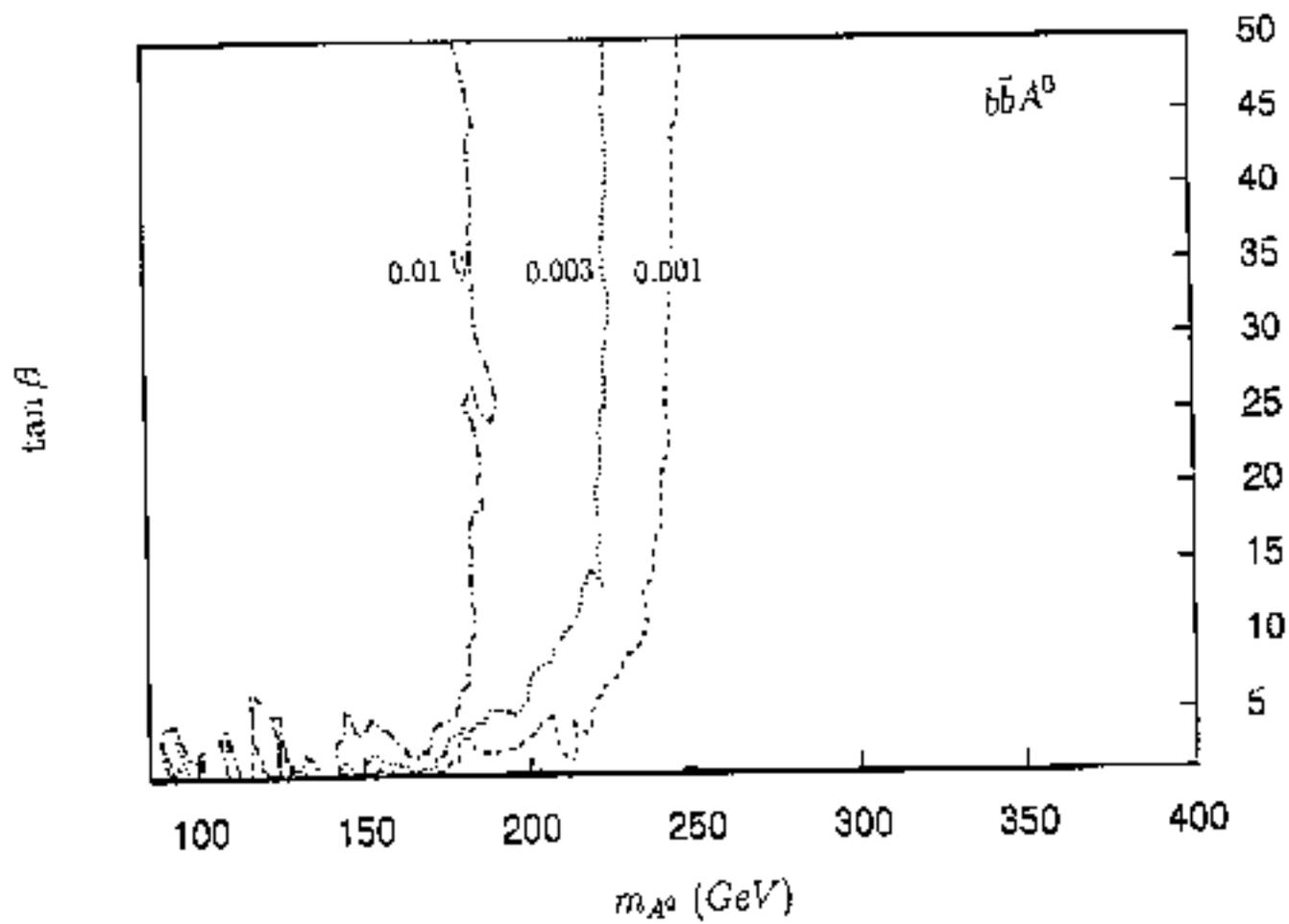
\noindent {\bf Fig. 7} Same as in Fig. 4, but for
$e^{+}e^{-}\rightarrow b \bar b A^0$.

\end{document}